\documentclass[hyper]{JHEP3} 

\usepackage{epsfig}

\def\npb#1{Nucl.~Phys.~{\bf B#1}}

\def\mt{{\ifmmode M^{eff}_T\else $M^{eff}_T$\fi}}

\def\e{\epsilon}

\def\ra{\rangle}

\def\e3{$\epsilon_3$}

\def\ch2{$\chi^2$}

\def\co#1{{\ifmmode{\cal O}_{#1}\else${\cal O}_{#1}$\fi}}

\newdimen\unit
\def\point#1 #2 #3{\vbox to0pt{\kern-#2\unit
  \hbox{\kern#1\unit#3}\vss}
 \nointerlineskip}

\newcommand{\be}{\begin{equation}}
\newcommand{\ee}{\end{equation}}
\newcommand{\bea}{\begin{eqnarray}}
\newcommand{\eea}{\end{eqnarray}}

\newcommand{\msbar}{\overline{\mbox{MS}}}

\newcount\hour
\newcount\minute
\newtoks\amorpm
\hour=\time\divide\hour by60
\minute=\time{\multiply\hour by60 \global\advance\minute by-
\hour}
\edef\standardtime{{\ifnum\hour<12 \global\amorpm={am}%
    \else\global\amorpm={pm}\advance\hour by-12 \fi
    \ifnum\hour=0 \hour=12 \fi
    \number\hour:\ifnum\minute<100\fi\number\minute\the\amorpm}}
\edef\militarytime{\number\hour:\ifnum\minute<100\fi\number\minute}
\def\bold#1{\setbox0=\hbox{$#1$}%
     \kern-.025em\copy0\kern-\wd0
     \kern.05em\copy0\kern-\wd0
     \kern-.025em\raise.0433em\box0 }

\newcommand{\newc}{\newcommand}
\newc\eg{{\it {e.g.}}}  \newc\etal{{\it {et al.}}} \newc\ie{{\it i.e.}}
\newc\etc{{\it {etc}}}

\newcommand\lsim{\mathrel{\rlap{\lower4pt\hbox{\hskip1pt$\sim$}}
    \raise1pt\hbox{$<$}}}
\newcommand\gsim{\mathrel{\rlap{\lower4pt\hbox{\hskip1pt$\sim$}}
    \raise1pt\hbox{$>$}}}

\newc{\mhalf}{m_{1/2}}      \newc{\mzero}{m_0}

\newc{\tanb}{\tan\beta}
\newc{\azero}{A_0}
\newc{\at}{A_t} \newc{\ab}{A_b} \newc{\atau}{A_\tau}

\newc{\bmu}{B\mu}           \newc{\sgn}{{\rm sgn}}
\newc{\mone}{M_1}           \newc{\mtwo}{M_2}

\newc{\charone}{\chi_1^\pm} \newc{\mcharone}{m_{\chi_1^\pm}}

\newc{\hl}{h}               \newc{\mhl}{m_{\hl}}
\newc{\hh}{H}               \newc{\mhh}{m_{\hh}}
\newc{\ha}{A}               \newc{\mha}{m_{\ha}}
\newc{\hc}{H^{\pm}}         \newc{\mhc}{m_{\hc}}

\newc{\qzero}{Q_0}          \newc{\qstop}{Q_{\widetilde t}}

\newc{\amu}{a_{\mu}}        \newc{\amususy}{a_{\mu}^{\rm SUSY}}
\newc{\amuexpt}{a_{\mu}^{\rm expt}}        \newc{\amusm}{a_{\mu}^{\rm SM}}
\newc{\deltaamususy}{\Delta a_{\mu}^{\rm SUSY}}
\newc\gmtwo{(g-2)_{\mu}} \newc\deltaamu{\Delta a_{\mu}}

\newc{\yt}{h_t} \newc{\yb}{h_b} \newc{\ytau}{h_{\tau}}
\newc{\mtpole}{m_t^{\rm pole}} \newc{\mbpole}{m_b^{\rm pole}}
\newc{\mtaupole}{m_{\tau}^{\rm pole}}

\newc{\mtmtsmmsbar}{m_t(m_t)^{\msbar}_{{\rm SM}}}
\newc{\mtmtsmdrbar}{m_t(m_t)^{\drbar}_{{\rm SM}}}
\newc{\mtmtmssmdrbar}{m_t(m_t)^{\drbar}_{{\rm SUSY}}}

\newc{\mbmbsmmsbar}{m_b(m_b)^{\msbar}_{{\rm SM}}}
\newc{\mbmzsmmsbar}{m_b(\mz)^{\msbar}_{{\rm SM}}}
\newc{\mbmzsmdrbar}{m_b(\mz)^{\drbar}_{{\rm SM}}}
\newc{\mbmzmssmdrbar}{m_b(\mz)^{\drbar}_{{\rm SUSY}}}
\newc{\mtaumzsmmsbar}{m_{\tau}(\mz)^{\msbar}_{{\rm SM}}}
\newc{\mtaumzsmdrbar}{m_{\tau}(\mz)^{\drbar}_{{\rm SM}}}
\newc{\mtaumzmssmdrbar}{m_{\tau}(\mz)^{\drbar}_{{\rm SUSY}}}

\newc{\mgut}{M_{\rm GUT}}

\newc{\mplanck}{M_{\rm P}}      \newc{\mpl}{M_{\rm Pl}}
\newc{\msusy}{M_{\rm SUSY}}      \newc{\ms}{M_{\rm S}}

\newc{\jxf}{J({\xf})}
\newc{\jxfexact}{J_{\rm exact}({\xf})}  \newc{\jxfexp}{J_{\rm exp}({\xf})}
\newc{\VEV}[1]{\langle #1 \rangle}

\newc{\xf}{x_f}
\newc\vrel{v_{\rm rel}}
\newcommand\mchi{m_{\chi}}              

\newc\sell{{\widetilde e}_L}      \newc\msell{m_{\sell}}
\newc\selr{{\widetilde e}_R}      \newc\mselr{m_{\selr}}

\newc\snue{{\widetilde \nu}_e}      \newc\msnue{m_{\snue}}
\newc\snutau{{\widetilde \nu}_\tau}      \newc\msnutau{m_{\snutau}}

\newc\supl{{\widetilde u}_L}      \newc\msupl{m_{\supl}}
\newc\supr{{\widetilde u}_R}      \newc\msupr{m_{\supr}}
\newc\sdl{{\widetilde d}_L}      \newc\msdl{m_{\sdl}}
\newc\sdr{{\widetilde d}_R}      \newc\msdr{m_{\sdr}}

\newc\hpm{H^\pm} \newc\hp{H^+} \newc\hm{H^-}
\newc\sfermion{\tilde f}  \newc\msfermion{m_{\sfermion}}

\newc\second{{\rm sec}}

\newc\alphas{\alpha_s}

\newc\alphaem{\alpha_{em}}

\newc{\gstar}{g_\ast}           \newc{\gsstar}{g_{s\ast}}
\newc{\geff}{g_{\rm eff}}

\newcommand\mz{m_{Z}}

\newc{\sthw}{\sin\theta_W}              \newc{\cthw}{\cos\theta_W}
\newc{\bino}{\widetilde B}              \newc{\wino}{\widetilde W_3^0}
\newc{\higgsinob}{{\widetilde H}^0_b}   \newc{\higgsinot}{{\widetilde H}^0_t}

\newc{\abund}{\Omega h^2}
\newc{\abundchi}{\Omega_\chi h^2}
\newc{\abundcdm}{\Omega_{CDM} h^2}
\newc{\omegam}{\Omega_{M}}       \newc{\abundm}{\Omega_{M} h^2}
\newc{\omegab}{\Omega_{b}}       \newc{\abundb}{\Omega_{b} h^2}
\newc{\omegacdm}{\Omega_{CDM}}
\newc{\omegatot}{\Omega_{TOT}}

\newc{\rhocrit}{\rho_{crit}}
\newc{\rhochi}{\rho_{\chi}}

\newc\br{\mbox{BR}}

\newc{\beq}{\begin{equation}}
\newc{\eeq}{\end{equation}}

\newcommand\vs{{\it {vs.}}}

\newc\stoponetwo{{\widetilde t}_{1,2}}
\newc\sbotonetwo{{\widetilde b}_{1,2}}
\newc\stauonetwo{{\widetilde \tau}_{1,2}}

\newc\bsgamma{b\ra s \gamma }
\newc\brbsgamma{\br( B\rightarrow X_s \gamma )}

\newc{\sigsip}{\sigma^{SI}_{p}} \newc{\sigsin}{\sigma^{SI}_{n}}
\newc{\sigsdp}{\sigma^{SD}_{p}} \newc{\sigsdn}{\sigma^{SD}_{n}}
\newc{\sigsiA}{\sigma^{SI}_{A}}

\long\def\begincomment#1\endcomment{%
        \begingroup\sf\baselineskip12pt#1\endgroup}

\title{Dark Matter And {\boldmath $B_s\rightarrow \mu^+\ \mu^-$}\\
With Minimal {\boldmath $SO_{10}$} Soft SUSY Breaking}

\author{Radovan Derm\' \i \v sek\\
        Davis Institute for High Energy Physics,\\
    University of California, Davis, CA 95616, USA\\
    E-mail: \email{dermisek@physics.ucdavis.edu}}

\author{Stuart Raby\\
        Department of Physics, The Ohio State University, \\
174 W. 18th Ave., Columbus, Ohio  43210, USA; \\
On leave of absence, School of Natural Sciences, \\ Institute for
Advanced Study,  Princeton, NJ 08540, USA \\
        E-mail: \email{raby@pacific.mps.ohio-state.edu}}

\author{Leszek Roszkowski\\
        Department of Physics, Lancaster University,
        Lancaster LA1 4YB, England\\
        E-mail: \email{L.Roszkowski@lancaster.ac.uk}}

\author{Roberto Ruiz de Austri\\
        Physics Division, School of Technology,
        Aristotle University of Thessaloniki, \\
        GR - 540 06 Thessaloniki, Greece \\
        E-mail: \email{rruiz@gen.auth.gr}}

\abstract{CMSSM boundary conditions are usually used when
calculating cosmological dark matter densities. In this paper we
calculate the cosmological density of dark matter in the MSSM
using minimal $SO_{10}$ soft SUSY breaking boundary conditions.
These boundary conditions incorporate several attractive features:
they are consistent with $SO_{10}$ Yukawa unification, they result
in a ``natural" inverted scalar mass hierarchy and they reduce the
dimension 5 operator contribution to the proton decay rate.  With
regards to dark matter, on the other hand, this is to a large
extent an unexplored territory with large squark and slepton
masses $m_{16}$, large $A_0$ and small $ \{ \mu, M_{1/2} \} $.  We
find that in most regions of parameter space the cosmological
density of dark matter is considerably less than required by the
data.  However there is a well--defined, narrow region of
parameter space which provides the observed relic density of dark
matter, as well as a good fit to precision electroweak data,
including top, bottom and tau masses, and acceptable bounds on the
branching fraction of $B_s \rightarrow \mu^+\ \mu^-$. We present
predictions for Higgs and SUSY spectra, the dark matter detection
cross section and the branching ratio ${\rm BR}(B_s\rightarrow
\mu^+\ \mu^-)$ in this region of parameter space.}


\preprint{OHSTPY-HEP-T-03-002\\UCD-03-02}
\begin{document}


\section{ Introduction}\label{intro:sec}

The constrained minimal supersymmetric standard model
[CMSSM]~\cite{cmssm} is a well defined model for soft SUSY
breaking with five independent parameters given by $m_0, \
M_{1/2}, \ A_0, \ $ $ \tan\beta$ and $sign(\mu)$.   It has been
used extensively for benchmark points for collider searches, as
well as for astrophysical and dark matter analyses.  The economy
of parameters in this scheme makes it a useful tool for exploring
SUSY phenomena. However the CMSSM may miss regions of soft SUSY
breaking parameter space which give qualitatively different
predictions.   In this paper we consider an alternate scheme, the
minimal $SO_{10}$ supersymmetric model [MSO$_{10}$SM], which is
well motivated and opens up a qualitatively new region of
parameter space.

In the MSO$_{10}$SM there are 7 soft SUSY breaking parameters
$\mu,\; M_{1/2},\; A_0,\; \tan\beta$,  $m_{16}$ (a universal
squark and slepton mass), $m_{10}$ (a universal Higgs mass) and
$\Delta m_H^2$ (Higgs up/down mass splitting). Moreover the
parameters $A_0, \ m_{10}, \ m_{16}$ must satisfy the
constraints~\cite{bdr,Tobe:2003bc,Auto:2003ys} $A_0 \approx  - 2 \
m_{16}$, $m_{10} \approx \sqrt{2} \ m_{16}$, $m_{16} >  1.2 \;
{\rm TeV}$ with $\mu, \ M_{1/2} \ll  m_{16}$ and $\tan\beta
\approx 50$. Note, with these values of the soft SUSY breaking
parameters, we can explore SUSY phenomena with qualitatively
different behavior than in the CMSSM. This is mainly due to the
Higgs splitting ($\Delta m_H^2$) which, as is well
known~\cite{ewsb}, enables one to obtain electroweak symmetry
breaking with values $m_{16} \gg \mu, \ M_{1/2}$.  Also, radiative
EWSB with $\tan\beta \approx 50$ requires significantly less fine
tuning with Higgs mass splitting (see Rattazzi and
Sarid~\cite{ewsb}).  Furthermore, with 3 Higgs mass parameters
$\mu, \ m_{10}, $ and $\Delta m_H^2$ we find that the latter two
are strongly constrained by EWSB, once we fix the value of $\mu$,
which we treat as a free parameter. This is unlike the CMSSM where
$\mu$ is fixed by EWSB.   Also note that small changes in $\Delta
m_H^2$ lead to big changes in the CP odd Higgs mass
$m_{A}$~\cite{bdr}.

It is not at all obvious that the MSO$_{10}$SM region of soft SUSY
breaking parameter space is consistent with
cosmology~\cite{Tobe:2003bc,Auto:2003ys}.\footnote{ See also other
recent articles discussing Yukawa unification and dark
matter~\cite{Chattopadhyay:2001va,Pallis:2003jc}} The dark matter
candidate in this model is the lightest neutralino. However, since
the scalar masses of the first two families are of order $m_{16} >
1.2 \; {\rm TeV}$, and the third generation sfermions (except for
the stops) also tend to be heavy, the usually dominant
annihilation channels, for the neutralino LSP to light fermions
via $t$--channel sfermion exchange, are suppressed. On the other
hand, the process $\chi\chi\rightarrow f\bar f$ via $s$--channel
$A$ exchange becomes important. This is due to the enhanced CP odd
Higgs coupling to down--type fermions, which is proportional to
$\tan\beta$, and because, in contrast to heavy scalar exchange,
the process is not $p$--wave suppressed. In an earlier analysis,
our $\chi^2$--analysis favored a light CP odd Higgs mass
$m_{A}\sim100$~GeV~\cite{bdr}, although heavier $A$ were also
allowed. Such light $A$ are however disfavored for two reasons. In
order to provide efficient annihilation for the LSPs, one would be
squeezed into a rather low LSP mass region $\mchi\approx m_{A}/2$,
which would require extreme fine--tuning at best. In addition,
such low $m_{A}$ are anyway inconsistent with the current limits
on ${\rm BR}(B_s\rightarrow \mu^+\ \mu^-)$. In this analysis, we
vary the $A$ mass.

We study the cosmology of the MSO$_{10}$SM in this paper.
Obtaining the observed relic abundance of cold dark matter, which
along with other cosmological parameters has recently been
determined with an unprecedented accuracy~\cite{wmap0302}, will
provide a new important constraint on the model.  We also compute
the branching ratio for the process $B_s \rightarrow \mu^+ \
\mu^-$ due to $A$ exchange~\cite{bsmumu}. It is absolutely
essential to include this latter constraint in our analysis. Note,
the CDF bound ${\rm BR} (B_s \rightarrow \mu^+ \ \mu^-) < 2.6
\times 10^{-6}$~\cite{cdf}.  The cross section for the direct
detection of dark matter is also computed. In
section~\ref{sec:m10ssm} we define the MSO$_{10}$SM, describe its
virtues and outline the analysis.  In section~\ref{sec:cdm} we
compute the cosmological dark matter density and discuss our
results. Then in section~\ref{sec:predictions} we discuss our
predictions for underground dark matter searches and for collider
Higgs and SUSY searches.

\section{ Minimal ${\bf SO_{10}}$ SUSY Model -- MSO$_{10}$SM }
\label{sec:m10ssm}

\subsection{Framework}
\label{sec:framework}

Let us define the minimal $SO_{10}$ SUSY model.  Quarks and
leptons of one family reside in the $\bf 16$ dimensional
representation, while the two Higgs doublets of the MSSM reside in
one $\bf 10$ dimensional representation.   For the third
generation we assume the minimal Yukawa coupling term given by \be
{\bf \lambda \  16 \ 10 \ 16 }.  \ee  On the other hand, for the
first two generations and for their mixing with the third, we
assume a hierarchical mass matrix structure due to effective
higher dimensional operators.   Hence the third generation Yukawa
couplings satisfy   $\lambda_t = \lambda_b = \lambda_\tau =
\lambda_{\nu_\tau} = {\bf \lambda}$.

Soft SUSY breaking parameters are also consistent with $SO_{10}$
with \begin{itemize}  \item  a universal gaugino mass  $M_{1/2}$,
\item a universal squark and slepton mass
$m_{16}$,\footnote{$SO_{10}$ does not require all sfermions to have
the same mass.  This however may be enforced by non--abelian family
symmetries or possibly by the SUSY breaking mechanism.}
\item a
universal scalar Higgs mass $m_{10}$, \item and a universal A
parameter $A_0$.
\end{itemize}
In addition we have the soft SUSY breaking Higgs mass parameters
$\mu$ and $B \mu$.  $B \mu$ may, as in the CMSSM, be exchanged for
$\tan\beta$.  Note, not all of these parameters are independent.
Indeed, in order to fit the low energy electroweak data, including
the third generation fermion masses, it has been shown that $A_0,
\ m_{10}, \ m_{16}$ must satisfy the constraints~\cite{bdr}  \bea
&A_0& \approx - 2 \ m_{16}  \nonumber \\
&m_{10}& \approx \sqrt{2} \ m_{16} \nonumber \\
&m_{16}& > 1.2 \; {\rm TeV} \nonumber \\
&\mu,& \ M_{1/2} \ll m_{16}
\label{eq:constraint} \eea with \be \tan\beta \approx  50.
\label{eq:tanbeta} \ee This result has been confirmed in two
recent analyses~\cite{Tobe:2003bc,Auto:2003ys}.\footnote{Note,
different regions of parameter space consistent with Yukawa
unification have also been discussed
in~\cite{Tobe:2003bc,Auto:2003ys,Balazs:2003mm}} The first
property (Eqn.~(\ref{eq:constraint})) is necessary to fit the top,
bottom and $\tau$ masses, in addition to the precision electroweak
data~\cite{bdr,Tobe:2003bc,Auto:2003ys}. The second property
(Eqn.~(\ref{eq:tanbeta})) is a consequence of third generation
Yukawa unification, since $m_t(m_t)/m_b(m_t) \sim \tan\beta$.

One loop threshold corrections at the GUT scale lead to two
significant parameters we treat as free parameters, although they
are calculable in any GUT.   The first is a correction to gauge
coupling unification given by  \be \epsilon_3 \equiv
\left[\alpha_3(M_G) - \tilde \alpha_G\right]/\tilde \alpha_G \ee
where the GUT scale $M_G$ is defined as the scale where
$\alpha_1(M_G) = \alpha_2(M_G) \equiv \tilde \alpha_G$.  The
second is a Higgs splitting mass parameter defined by \be \Delta
m_H^2 \equiv (m_{H_d}^2 - m_{H_u}^2)/2 m_{10}^2 . \ee In order to
fit the low energy data we find $\epsilon_3 \approx - 4\%$ and
$\Delta m_H^2 \approx 13 \%$~\cite{bdr}. The largest corrections
to $\epsilon_3$ come from the Higgs and $SO_{10}$ breaking
sectors,  while the correction to $\Delta m_H^2$ is predominantly
due to the right--handed $\tau$ neutrino. For $M_{\bar \nu_\tau}
\approx 10^{13-14}$ GeV (appropriate for a light $\tau$ neutrino
mass $\approx 0.06$ eV) we obtain $\Delta m_H^2 \approx 10 - 7
\%$.

Finally, as a bonus, these same values of soft SUSY breaking
parameters, with $m_{16} \gg$ TeV, result in two very interesting
consequences.   Firstly, it ``naturally" produces an inverted
scalar mass hierarchy [ISMH]~\cite{scrunching}. With an ISMH
squarks and sleptons of the first two generations obtain mass of
order $m_{16}$ at $M_Z$. The stop, sbottom, and stau, on the other
hand, have mass less than a TeV. An ISMH has two virtues.
\begin{enumerate}
\item It preserves ``naturalness" (for values of $m_{16}$ which
are not too large), since only the third generation squarks and
sleptons couple strongly to the Higgs.

\item It ameliorates the SUSY CP and flavor problems, since these
constraints on CP violating angles or flavor violating squark and
slepton masses are strongest for the first two generations, yet
they are suppressed as $1/m_{16}^{2}$.  For $m_{16} > $ a few TeV,
these constraints are weakened~\cite{masieroetal}.
\end{enumerate}

Secondly, Super--Kamiokande bounds on $\tau(p \rightarrow K^+ \bar
\nu) \ > 1.9 \times 10^{33}$ yrs.~\cite{superk}  constrain the
contribution of dimension 5 baryon and lepton number violating
operators.   These are however minimized with $\mu, \ M_{1/2} \ll
m_{16}$~\cite{pdecay}.

\subsection{Analysis}

We use a top--down approach with a global \ch2
analysis~\cite{chi2}. The input parameters are defined by boundary
conditions at the GUT scale. The 11 input parameters at $M_G$ are
given by --- three gauge parameters $M_G, \; \alpha_G(M_G),$
$\epsilon_3$; the Yukawa coupling $\lambda$, and 7 soft SUSY
breaking parameters $\mu,\; M_{1/2},\; A_0,\; \tan\beta$, $\;
m_{16}^2, \; m_{10}^2, \Delta m_H^2$.  These are fit in a global
$\chi^2$ analysis defined in terms of physical low energy
observables.   Note we keep three parameters $ ( m_{16}, \ \mu, \
M_{1/2} )$ fixed; while minimizing $\chi^2$ with the remaining 8
parameters.  Below we will plot $\chi^2$ contours as a function of
$\mu, \ M_{1/2}$ for different values of $m_{16}$.  We use two
(one)~loop renormalization group [RG] running for dimensionless
(dimensionful) parameters from $M_G$ to $M_Z$.\footnote{Note, we
have checked that switching to 2 loop RGEs for dimensionful
parameters can be compensated for by small changes in the GUT
scale parameters, without significant changes in the low energy
results.}  We require electroweak symmetry breaking using an
improved Higgs potential, including $m_t^4$ and $m_b^4$
corrections in an effective 2 Higgs doublet model below $M_{SUSY}
= \sqrt{\frac{1}{2} (m_{\tilde t_1}^2 + m_{\tilde
t_2}^2)}$~\cite{carenaetal}.

The $\chi^2$ function includes 9 observables; 6 precision
electroweak data $\alpha_{EM},$ $G_\mu,$ $\alpha_s(M_Z),$ $M_Z, \;
M_W,$ $\rho_{NEW}$ and the 3 fermion masses $M_{top},\;  m_b(m_b),
\; M_\tau$.   In our analysis we fit the central
values~\cite{pdg2000}: $M_Z = 91.188$ GeV, $M_W = 80.419$ GeV,
$G_{\mu}\times 10^5  = 1.1664$ GeV$^{-2}$, $\alpha_{EM}^{-1} =
137.04,$ $M_{\tau} = 1.7770$ GeV with 0.1\% numerical
uncertainties; and the following with the experimental uncertainty
in parentheses: $\alpha_s(M_Z) = 0.1180\; (0.0020),$
$\rho_{new}\times 10^3 = -0.200\; (1.1)$~\cite{rhonew}, $M_t =
174.3\; (5.1)$ GeV, $m_b(m_b) = 4.20\; (0.20)$ GeV.\footnote{Note
we take a conservative error for $m_b(m_b)$~\cite{pdg2000} in view
of recent claims to much smaller error bars~\cite{Beneke:1999fe}.}
We include the complete one loop threshold corrections at $M_Z$ to
all observables.  In addition we use one loop QED and three loop
QCD RG running below $M_Z$.

The output of this analysis is a set of weak scale squark,
slepton, gaugino and Higgs masses.  With regards to the calculated
Higgs and sparticle masses, the neutral Higgs masses $h,\; H, \;
A$ are pole masses calculated with the leading top, bottom, stop,
sbottom loop contributions; while all other sparticle masses are
running masses. This output is then used to compute the
cosmological dark matter density of the lightest neutralino which
is the LSP.  The dark matter analysis is discussed in more detail
in section~\ref{sec:cdm}.

Using $\chi^2$ penalties\footnote{In order to constrain the values
of some physical observables in our $\chi^2$ analysis, such as
$m_{\tilde t_1}$ or $m_{A}$, we add a significant contribution to
the $\chi^2$ function for values of these observables outside the
desired range. We refer to this additional contribution as a
$\chi^2$ penalty.   Minimization of $\chi^2$ with Minuit, then
pushes the fits to the desired range.   Of course the $\chi^2$
penalties then vanish.} we apply two additional constraints:

\begin{itemize}
\item  $m_{\tilde t_1} \geq 300$  GeV \item  $m_{A}$ {\bf fixed}.
\end{itemize}

The first is chosen to be consistent with ${\rm BR} (B\rightarrow
X_s\gamma)$~\cite{bdr}.  Note, although we do calculate ${\rm BR}
(B\rightarrow X_s\gamma)$, we do not use it as a constraint in the
analysis. This is for two reasons --- 1) this decay mode depends
on 3--2 generation mixing which is model dependent and 2) it is
not difficult to fit ${\rm BR} (B\rightarrow X_s\gamma)$ for
values of $m_{\tilde t_1} \geq 300$ GeV.  Hence, in order to be
generally consistent with the measured value of ${\rm BR}
(B\rightarrow X_s\gamma)$, we impose $m_{\tilde t_1} \geq 300$
GeV.  With regards to the second constraint, since $\abundchi$ and
${\rm BR} (B_s \rightarrow \mu^+ \ \mu^-)$ are both sensitive to
the value of $m_{A}$, we fix it's value and present our results
for different values of $m_{A}$.\footnote{The calculation of ${\rm
BR} (B\rightarrow X_s\gamma)$ and ${\rm BR} (B_s \rightarrow \mu^+
\ \mu^-)$ requires a model for fermion mass matrices.  In the
absence of such a model we use the observed CKM matrix elements to
calculate these flavor violating branching ratios.}

Finally, we apply the experimental limits:
\begin{itemize}
\item
lower bound on the
lightest chargino mass $m_{\chi^+} > 104$~GeV,
\item
lower bound on the light Higgs mass  $m_h > 111$~GeV.
\end{itemize}
Note, because of the theoretical uncertainty in the calculation of
$m_h$ ($\sim3$~GeV), we conservatively impose $m_h>111$~GeV,
instead of the LEP bound for SM Higgs $m_h>114.4$~GeV.

\section{ Cosmological Dark Matter Density} \label{sec:cdm}

We compute the relic abundance $\abundchi$ of the lightest
neutralino using exact expressions for neutralino pair
annihilation into all allowed final--state channels, which are
valid both near and further away from resonances and
thresholds~\cite{nrr1+2}. We further treat the neutralino
coannihilation with the lightest chargino and next--to--lightest
neutralino~\cite{eg97} and with the lighter stau~\cite{nrr3} with
similar precision. We only neglect the neutralino coannihilation
with the stop which would only affect $\abundchi$ in the regions
of parameter space which are uninteresting for other reasons, as
we comment below. We solve the Boltzmann equation numerically as
in~\cite{darksusy} and compute $\abundchi$ with an error of a few
per cent, which is comparable with today's accuracy on the
observational side.  The latest determinations of cosmological
parameters~\cite{wmap0302} give $\abundm= 0.135^{+0.008}_{-0.009}$
for the total matter content and $\abundb= 0.0224\pm 0.0009$ for
the baryonic component. The difference, attributed to cold dark
matter (CDM), is then \beq \abundcdm= 0.113\pm0.009, \eeq which is
significantly narrower than previous ranges. We then apply two
constraints on the dark matter abundance:
\begin{itemize}
\item  the upper bound $\abundchi < 0.13$,
\item  $2\,\sigma$ preferred range $0.095< \abundchi < 0.13$.
\end{itemize}

\begin{figure}[t!]
\begin{center}
\begin{minipage}{5in}
\epsfig{file=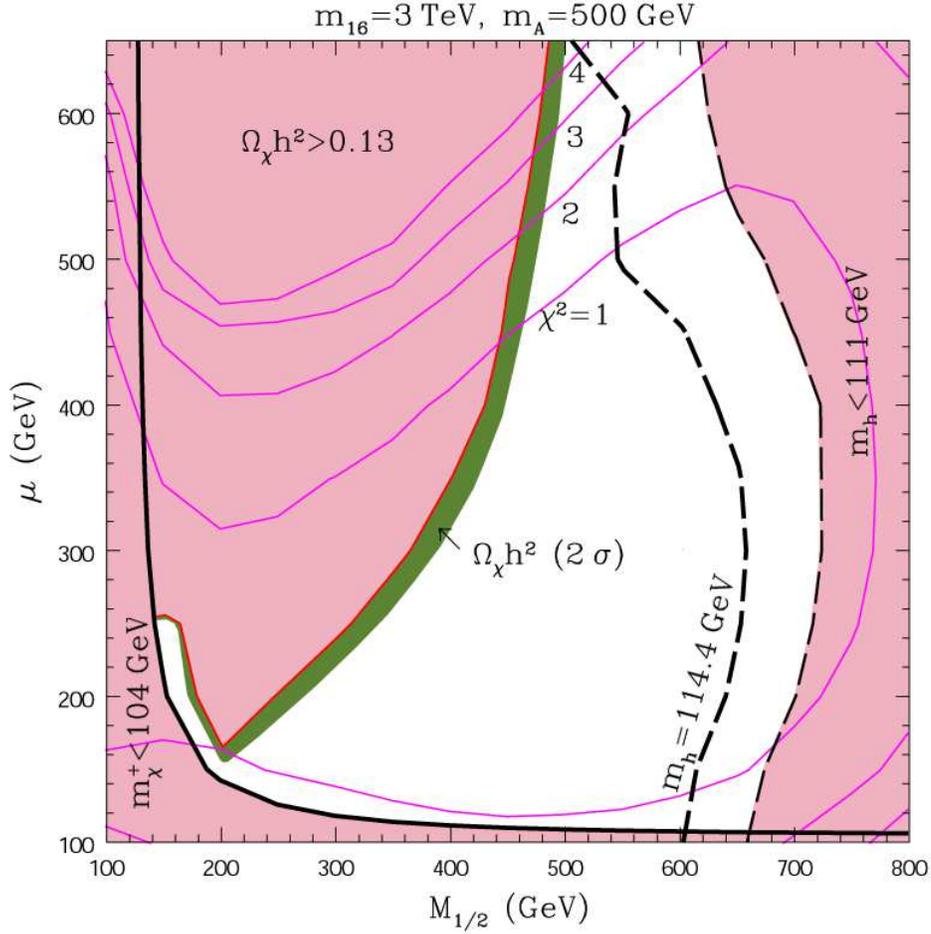,height=5in} 
\end{minipage}
\caption{\label{fig:3k500} {\small Contours of constant $\chi^2$
for $m_{16} = 3$ TeV and $m_{A} = 500$ GeV. The red regions are
excluded by $m_{\chi^+}<104$ GeV (below and to the left of a black
solid curve), $m_{h}<111$ GeV (on the right) and by
$\abundchi>0.13$. To the right of the black broken line one has
$m_{h}<114.4$ GeV. The green band corresponds to the preferred
$2\,\sigma$ range $0.095< \abundchi < 0.13$, while the white regions below
it correspond to $\abundchi<0.095$. }}
\end{center}
\end{figure}

\begin{figure}[t!]
\begin{center}
\begin{minipage}{6in}
\epsfig{file=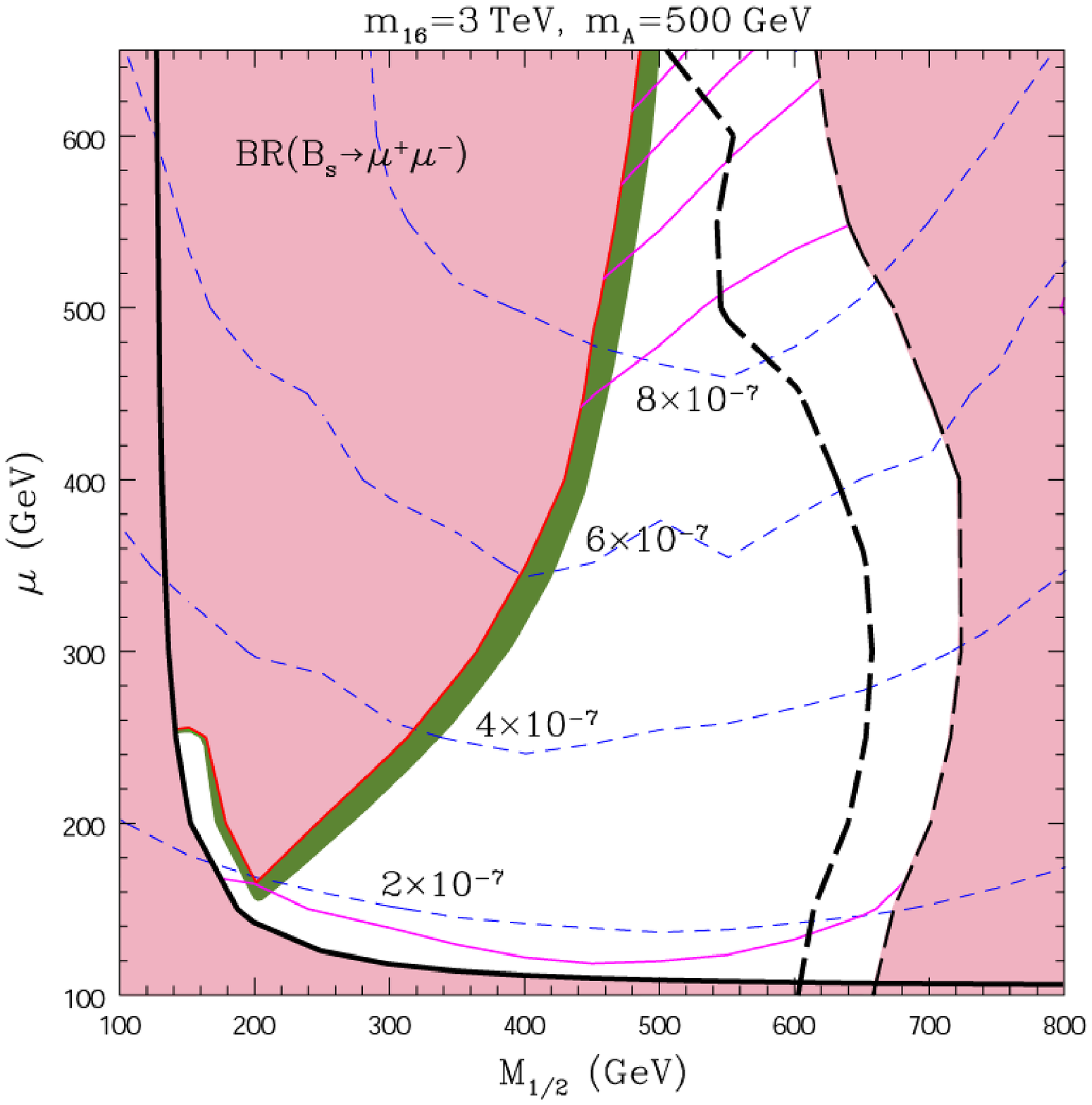,height=3in}
\hspace*{-0.15in}
\epsfig{file=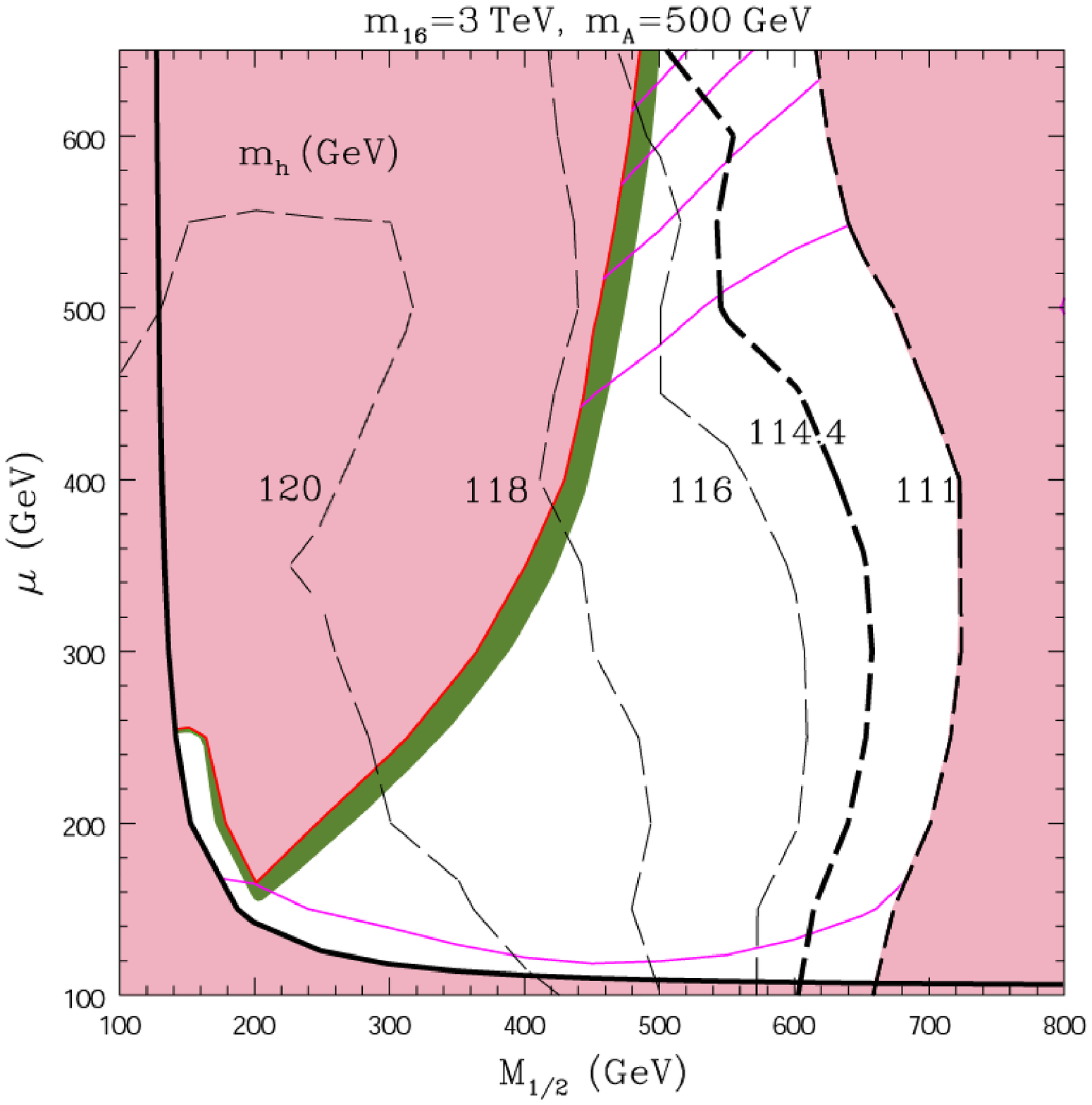,height=3in}
\end{minipage}
\end{center}
\vspace*{-.50in} 
\hspace*{-.70in}
\begin{center}
\begin{minipage}{6in}
\epsfig{file=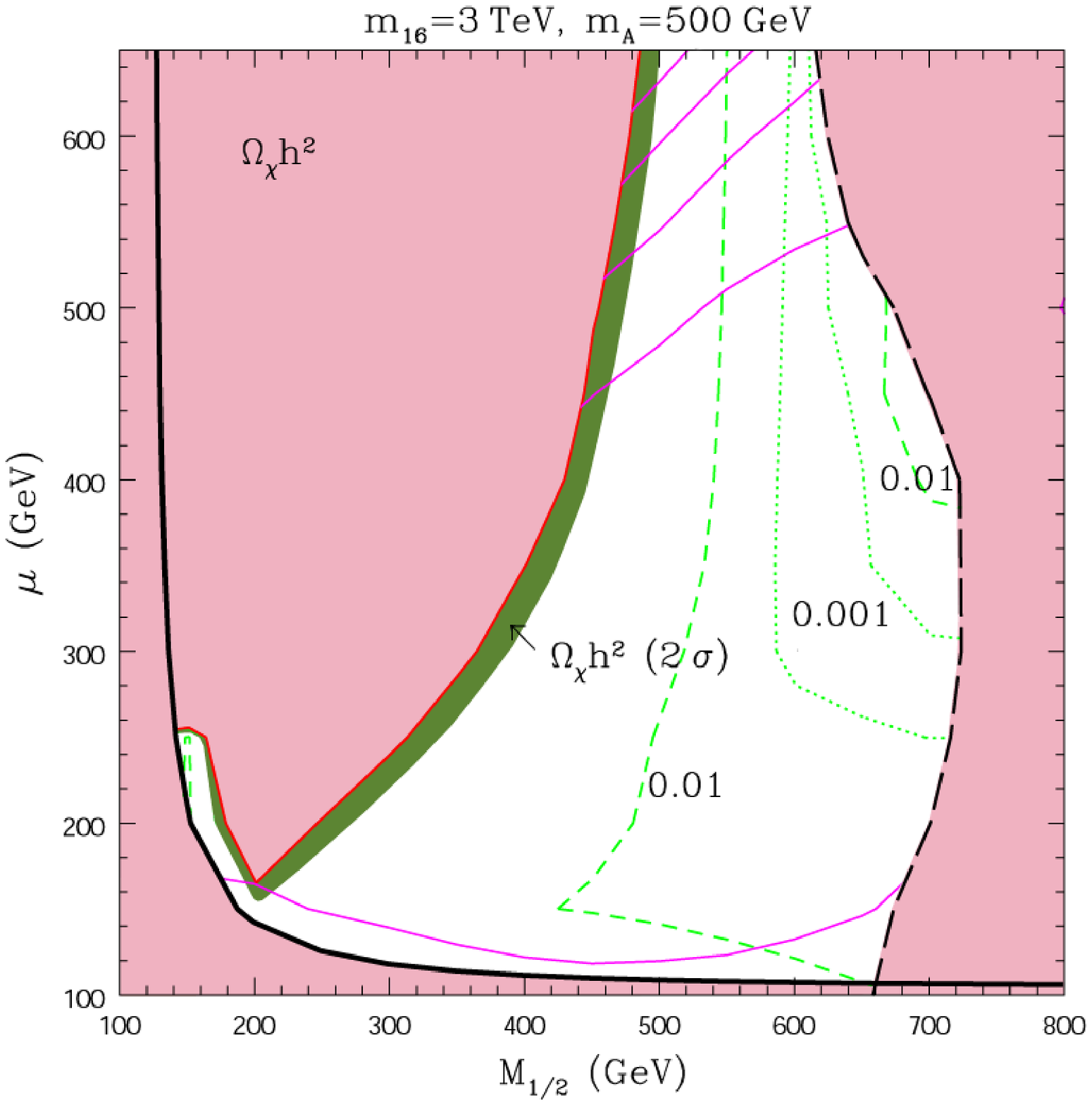,height=3in}
\hspace*{-0.15in}
\epsfig{file=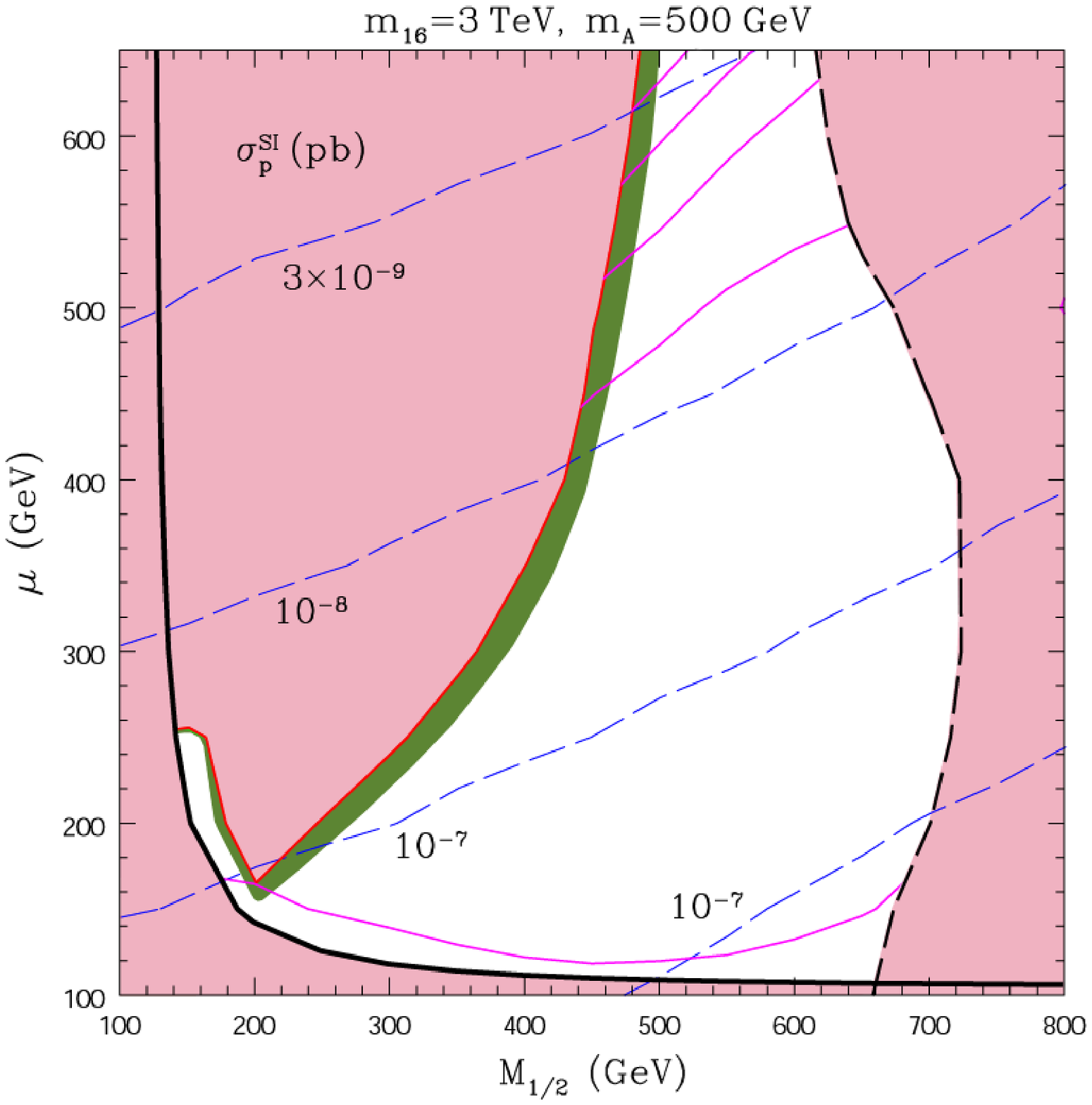,height=3in}
\end{minipage}
\caption{\label{fig:det3k500} {\small Same as Fig. 1 with contours
of constant ${\rm BR} (B_s\rightarrow \mu^+\ \mu^-)$ (upper left),
$m_h$ (upper right), $\abundchi$ (lower left) and $\sigsip$ (lower
right) for $m_{16} = 3$ TeV and $m_{A} = 500$ GeV. } }
\end{center}
\end{figure}

\begin{figure}[t!]
\begin{center}
\begin{minipage}{6in}
\epsfig{file=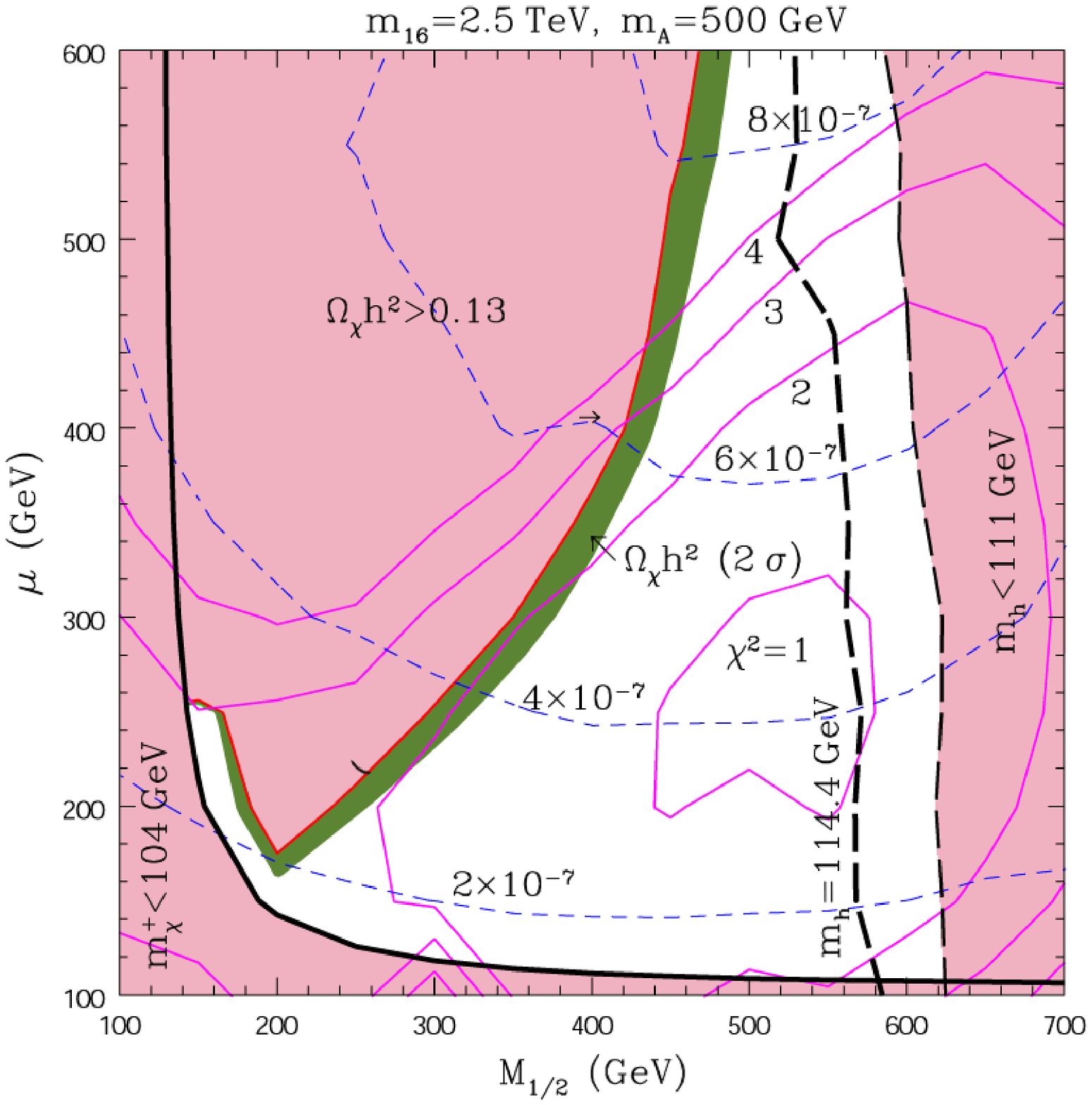,height=3in}
\hspace*{-0.15in}
\epsfig{file=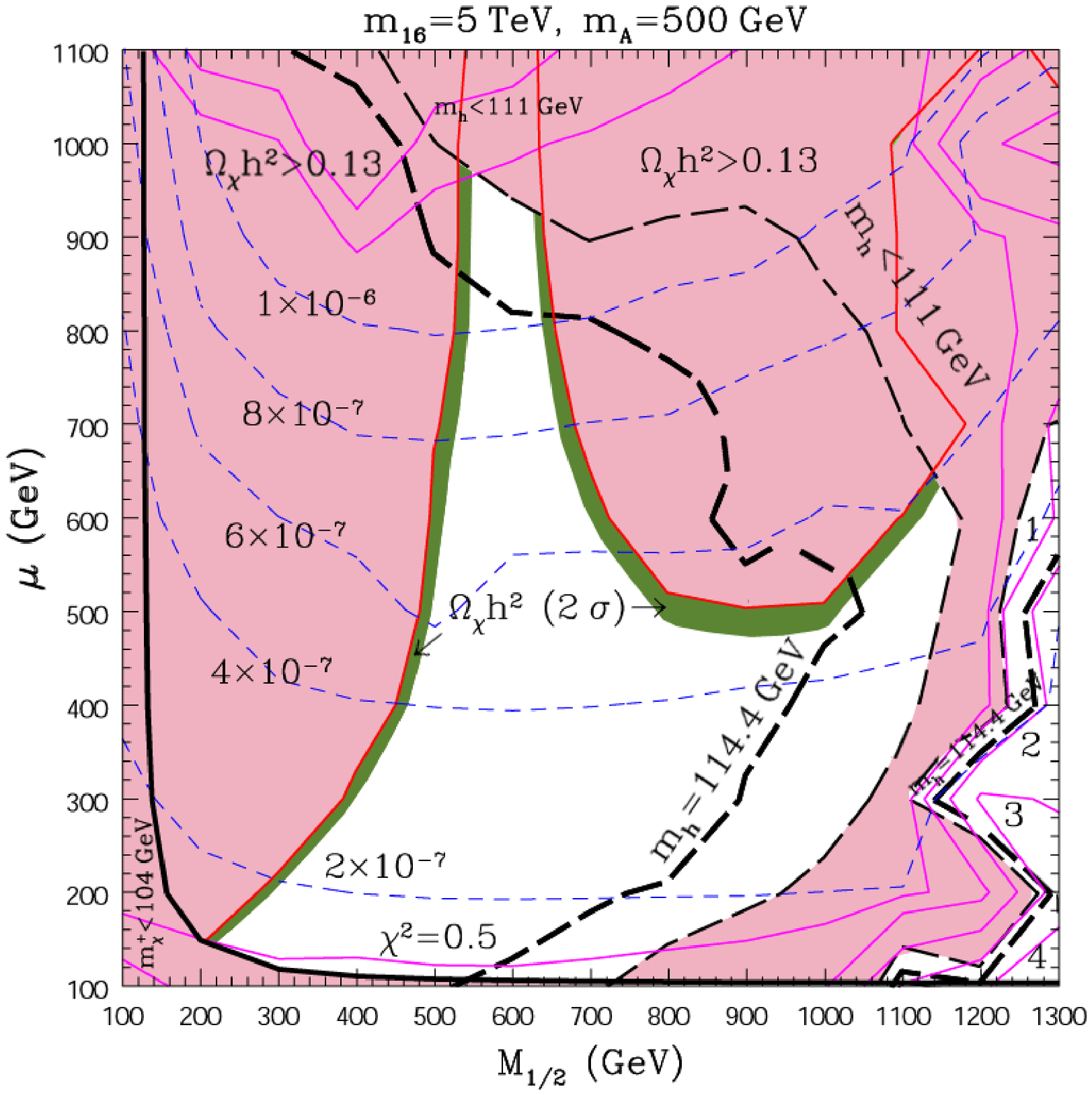,height=3in}
\end{minipage}
\end{center}
\vspace*{-.50in}
\hspace*{-.70in}
\begin{center}
\begin{minipage}{6in}
\epsfig{file=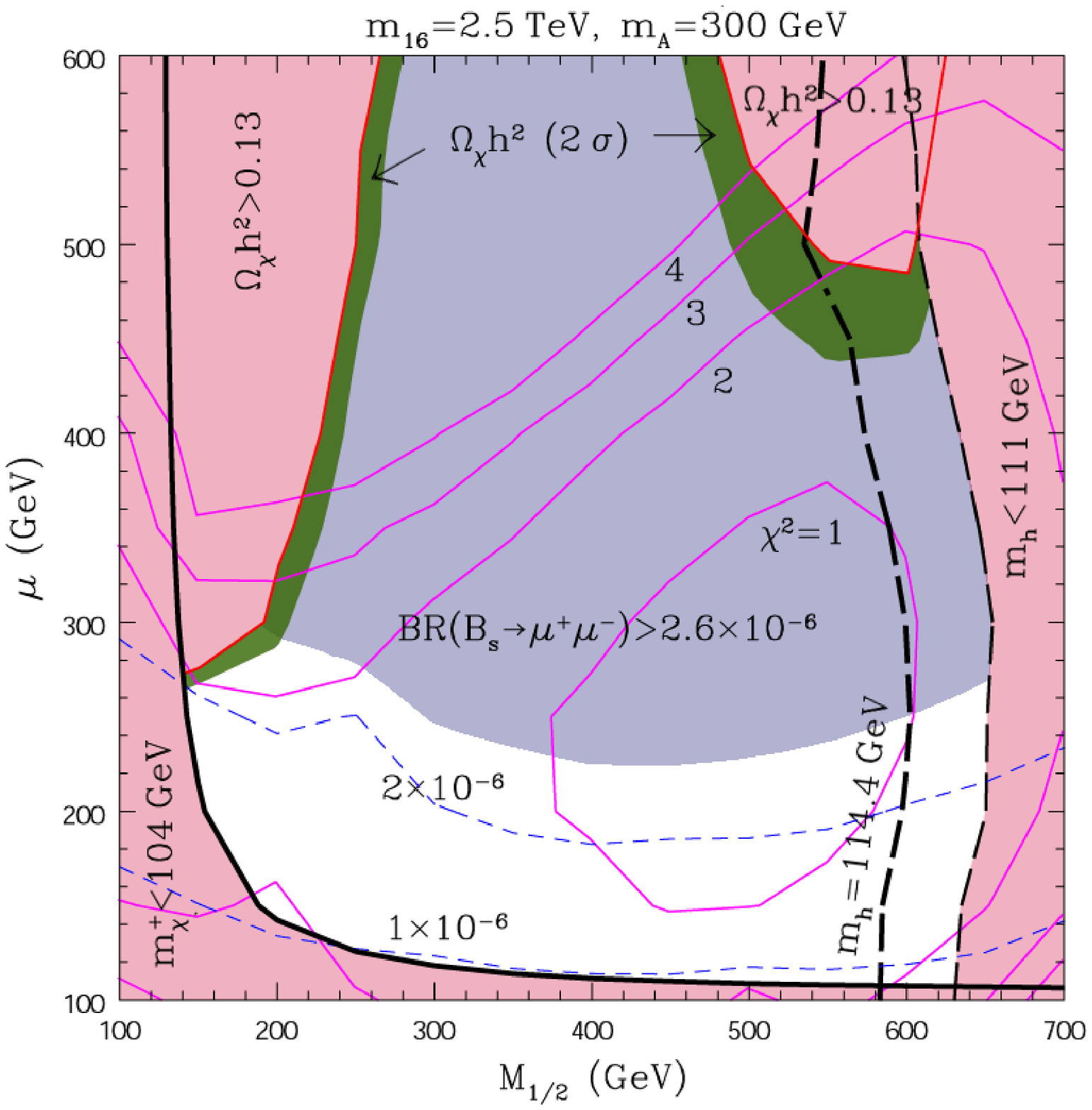,height=3in}
\hspace*{-0.15in}
\epsfig{file=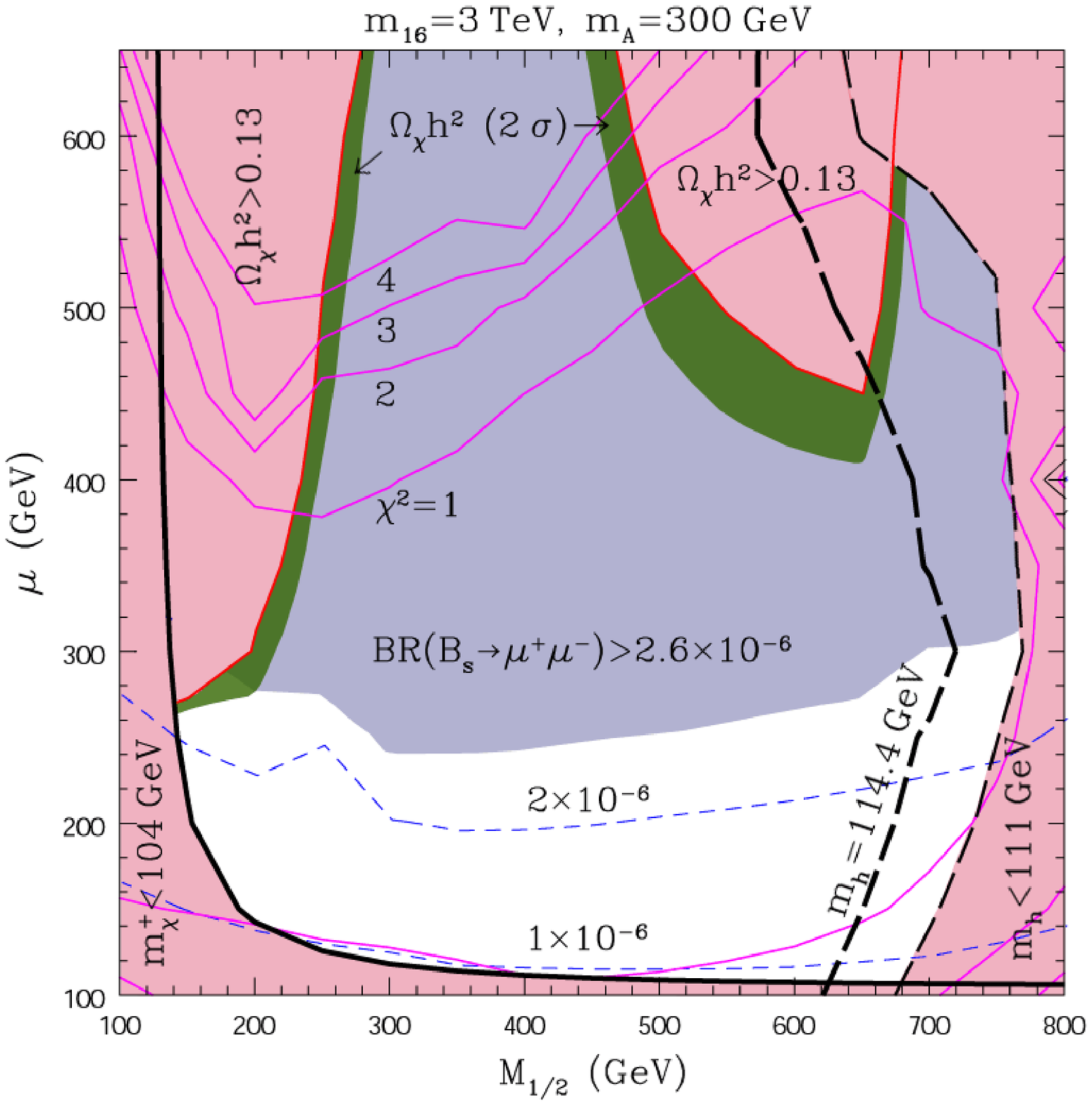,height=3in}
\end{minipage}
\caption{\label{fig:varm0ma} {\small The same as in
Fig.~\protect\ref{fig:3k500} but for $m_{16} = 2.5$ TeV and $m_{A} =
500$ GeV (upper left), $m_{16} = 5$ TeV and $m_{A} = 500$ GeV (upper
right), $m_{16} = 2.5$ TeV and $m_{A} = 300$ GeV (lower left) and
$m_{16} = 3$ TeV and $m_{A} = 300$ GeV (lower right). Also marked are
contours of constant ${\rm BR} (B_s\rightarrow \mu^+\ \mu^-)$. The
blue regions in the lower two panels are excluded by ${\rm BR}
(B_s\rightarrow \mu^+\ \mu^-)>2.6\times10^{-6}$.  Note the different
mass ranges for $M_{1/2}$ and $\mu$ in the different panels. }}
\end{center}
\end{figure}

In Figs.~\ref{fig:3k500}--\ref{fig:varm0ma} we present our results
for different values of $m_{16}$ and $m_{A}$ in the $\mu, \
M_{1/2}$ plane.    In particular in Fig.~\ref{fig:3k500} we
present, for $m_{16} = 3$ TeV and $m_{A} = 500$ GeV, the (magenta)
lines of constant $\chi^2$ with the cosmologically preferred dark
matter region (shaded green) satisfying  $0.095 < \abundchi <
0.13$.  We find significant regions of parameter space which gives
$\chi^2 \leq 2$,  $\abundchi$ as above, and satisfies all other
phenomenological constraints. In addition we have shaded (light
red) the regions excluded by collider limits and by $\abundchi >
0.13$.\footnote{A similar analysis was performed in the recent
paper~\cite{Auto:2003ys}.   However, they were not able to find
acceptable cosmological solutions.  It appears that they did not
find any acceptable solutions because they find Yukawa unification
only for $M_{1/2} \sim 100$ GeV and large $\mu \sim 300$ GeV. In
this region we probably would not find acceptable solutions for
$\abundchi$, no matter what value we take for $m_A$.}

In Fig.~\ref{fig:det3k500} we present a more detailed analysis of
the same $m_{16} = 3$ TeV, $m_{A} = 500$ GeV case given in Fig.
\ref{fig:3k500}.   We now include lines of constant ${\rm BR} (B_s
\rightarrow \mu^+ \ \mu^-)$ (upper left), $m_h$ (upper right),
$\abundchi$ (lower left), and $\sigsip$ (lower right). $\sigsip$
is the spin independent neutralino dark matter cross--section
relevant for direct dark matter searches. We now consider each one
of these features further.

Recall, the branching ratio ${\rm BR} (B_s \rightarrow \mu^+ \
\mu^-)$ is sensitive to the value of the CP odd Higgs mass
$m_{A}$~\cite{bsmumu}.  For $m_{A} = 500$ GeV, the branching ratio
satisfies $2 \times 10^{-7} < {\rm BR} (B_s \rightarrow \mu^+ \
\mu^-) < 8 \times 10^{-7}$ for acceptable values of $\abundchi$
and $\chi^2 < 2$. In a recent analysis it has been shown that,
with an integrated luminosity of 15 fb$^{-1}$, CDF can discover
this process if ${\rm BR} (B_s\rightarrow \mu^+\ \mu^-) > 1.2
\times 10^{-8}$~\cite{Arnowitt:2002cq}. Hence most acceptable
regions of parameter space lead to observable rates for ${\rm BR}
(B_s\rightarrow \mu^+\ \mu^-)$.

In Fig.~\ref{fig:det3k500} (upper right) we see that the light
Higgs mass increases as $M_{1/2}$ decreases.  In the acceptable
regions of parameter space we find $116 < m_h < 121$ GeV.   The
value of the light Higgs mass is however fairly insensitive to
$m_{16}$.

The cosmological relic abundance of the neutralino $\abundchi$
(Fig.~\ref{fig:det3k500} (lower left)) is primarily determined by
the direct s--channel pair--annihilation into SM fermion pairs
through the CP odd Higgs. Since all the sfermions are very heavy,
their contribution to reducing the neutralino number density is
strongly suppressed. In contrast, because of the coupling $A
b{\bar b}\propto \tan\beta$ (and similarly for the $\tau$'s), the
$A$--resonance is effective and broad. Near $\mchi\approx m_{A}/2$
it reduces $\abundchi$ down to allowed but uninterestingly small
values $\ll 0.1$. As one moves away from the resonance,
$\abundchi$ grows, reaches the preferred range $0.095 < \abundchi
< 0.13$, before becoming too large $\abundchi >
0.13$.\footnote{Note that at one loop we have $M_1(M_Z) = M_{1/2}
* \alpha_1(M_Z)/\alpha_G$ so $M_1(M_Z) \approx 0.4 M_{1/2}$. For
bino--like neutralino (which is true for larger $\mu$), we thus
have $\mchi \approx 0.4 M_{1/2}$.  Hence for s--channel
annihilation we have $m_{A} \approx 2 m_{\chi} \approx 0.8 \
M_{1/2}$ or $M_{1/2}\approx (5/4)\,m_{A}$ for the position of the
``peak suppression."}  (A similar, but much more narrow resonance
due to $h^0$ is also present at $M_{1/2}\approx 150$~GeV and small
$\mu$.) When $\mchi\gsim m_t$ ($M_{1/2}\gsim420$~GeV) and the
stops are not too heavy, the LSP pairs annihilate to $t\,{\bar
t}$--pairs.  In the region of large $M_{1/2}$, often where $m_h$
is already too low, two additional channels become effective.
First, in this region the neutralino becomes almost mass
degenerate with the lighter stau which leads to reducing
$\abundchi$ through coannihilation. Second, if $m_{A}$ is not too
large, neutralino pair--annihilation into Higgs boson pairs $AA$
and $HH$ opens up. Finally, at $\mu\ll M_{1/2}$, the relic
abundance is strongly reduced due to the increasing higgsino
component of the LSP.

Finally, the spin independent neutralino cross--section $\sigsip$
in the lower right window of Fig.~\ref{fig:det3k500} is
predominantly determined by the contribution of the heavy CP even
scalar $t$--channel exchange to both tree--level and one--loop
diagrams. Note that in the preferred region of $\chi^2 < 2$ and
$0.095<\abundchi<0.13$ we find $ 10^{-9}~{\rm pb} \lsim \sigsip \lsim
10^{-7}~{\rm pb}$. We will comment further on our predictions for
$\sigsip$ below.

In Fig.~\ref{fig:varm0ma} we display the dependence of our
constraints, $\abundchi$ and ${\rm BR} (B_s\rightarrow \mu^+\
\mu^-)$ on $m_{16}$ and $m_{A}$.  By comparing the upper two
windows with Fig.~\ref{fig:3k500} we can see that, as $m_{16}$
increases, the region with $\chi^2 < 2$ rapidly grows.  Note, the
dominant pull in $\chi^2$ is due to the bottom quark mass. In
order to fit the data, the total SUSY corrections to $m_b(m_b)$
must be of order $-(2 - 4) \%$~\cite{bdr}.   In addition there are
three dominant contributions to these SUSY corrections,  a gluino
loop contribution  $\propto \ \alpha_3 \ \mu \ M_{\tilde g} \
\tan\beta/m_{\tilde b_1}^2$, a chargino loop contribution $\propto
\ \lambda_t^2 \ \mu \ A_t \ \tan\beta/m_{\tilde t_1}^2$, and a
term $\propto \ \log M_{SUSY}^2$. When $m_{16}$ increases (with
$M_{1/2}$ fixed) the parameter $A_t$ becomes more negative, since
$A_0 \approx - 2 \ m_{16}$ and $A_t \approx - 3 \ M_{1/2} +
\epsilon \ A_0$ where $\epsilon \ll 1$. Also, larger values of
$m_{16}$ permit a larger range for the ratio $m_{\tilde
b_1}/m_{\tilde t_1}$. Thus larger values of $m_{16}$ allows more
freedom in parameter space for fitting the data at both smaller or
larger values of $\mu, \; M_{1/2}$.

In the lower two windows in Fig.~\ref{fig:varm0ma} we consider two
regions with $m_{A} = 300$ GeV with $m_{16} = 2.5$ TeV (lower
left) and $m_{16} = 3$ TeV (lower right).   The blue regions are
excluded by the CDF bound ${\rm BR} (B_s\rightarrow \mu^+\ \mu^-)
< 2.6 \times 10^{-6}$~\cite{cdf}. Note for $m_{16} = 2.5$ TeV, the
region with $\chi^2 < 2$ does not overlap the region with
acceptable dark matter abundance (green shaded). However for
$m_{16} = 3$ TeV, $m_{A} = 300$ GeV,
Fig.~\protect\ref{fig:varm0ma} (lower right), we find a small
region with acceptable $\chi^2 < 1$ and $\abundchi$. Moreover the
branching ratio ${\rm BR} (B_s\rightarrow \mu^+\ \mu^-)$ is now
close to the CDF bound.  On the other hand for $m_{16} = 5$ TeV,
$m_{A} = 500$ GeV, in Fig.~\protect\ref{fig:varm0ma} (upper
right), a new region of parameter space consistent with all the
data now opens up with larger $\mu, \ M_{1/2}$.  This new region
becomes cosmologically allowed due to neutralino annihilation into
Higgs boson pairs $AA$ and $HH$ and due to coannihilation with the
lighter stau.

Hence we see that increasing $m_{A}$ has two effects.  It
suppresses the branching fraction ${\rm BR} (B_s \rightarrow \mu^+
\ \mu^-)$. At the same time it moves the s--channel neutralino
annihilation channel to larger values of $M_{1/2}$; hence
providing larger regions with $0.095 < \abundchi < 0.13$ (compare
Figs.~\protect\ref{fig:3k500} and \protect\ref{fig:varm0ma} (lower
right) or Figs.~\protect\ref{fig:varm0ma} (upper and lower left)).
In fact, the two (green) branches of the preferred range $0.095 <
\abundchi < 0.13$  correspond to the two sides (except for the
upper left window of Fig.~\protect\ref{fig:varm0ma} where just one
side is evident) of the wide $A$ resonance in the neutralino
pair--annihilation. On the other hand, increasing $m_{A}$ above
1~TeV or so would move the regions of preferred $\abundchi$ too
far to the right, in potential conflict with a lower bound on
$m_h$.

Finally, we comment on ${\rm BR} (B\rightarrow X_s\gamma)$. The
current experimental range~\cite{bsgamma-exp,or1} is ${\rm BR} (B
\rightarrow X_s\gamma)_{\rm expt}=(3.41 \pm0.36)\times10^{-4}$,
while the SM prediction, including full NLO QCD
corrections~\cite{gm01,bcmu02}, is ${\rm BR} (B \rightarrow
X_s\gamma)_{\rm SM}=(3.70 \pm0.30)\times10^{-4}$.  In computing
the SUSY contribution to $b\rightarrow s\gamma$ we further include
full LO and dominant NLO--level $\tan\beta$--enhanced
contributions~\cite{dgg00,cgnw00}. Conservatively allowing for the
SM+SUSY contribution to be in the range
$\left(3.41\pm0.67\right)\times 10^{-4}$~\cite{rrn1,or1} selects a
band $300~{\rm GeV}\lsim \mu \lsim 400~{\rm GeV}$ which slowly
decreases with increasing $M_{1/2}$. (This remains approximately
true for all the cases that we have analyzed except $m_{16} = 5$
TeV and $m_{A} = 500$ GeV where one finds a narrower range at
$\mu\lsim200$~GeV.)  However, ${\rm BR} (B \rightarrow X_s\gamma)$
is strongly sensitive to the 2--3 generation down--type squark
mixings~\cite{or1} which are model dependent and which we do not
include here.  In summary, the process is generally consistent
with the most preferred regions of $M_{1/2}$ and $\mu$ but we do
not use it here as a constraint, since it can be easily relaxed by
employing parameters which are less relevant for our analysis.

\begin{figure}[t!]
\begin{center}
\hspace*{0.80in}
\begin{minipage}{4.5in}
\epsfig{file=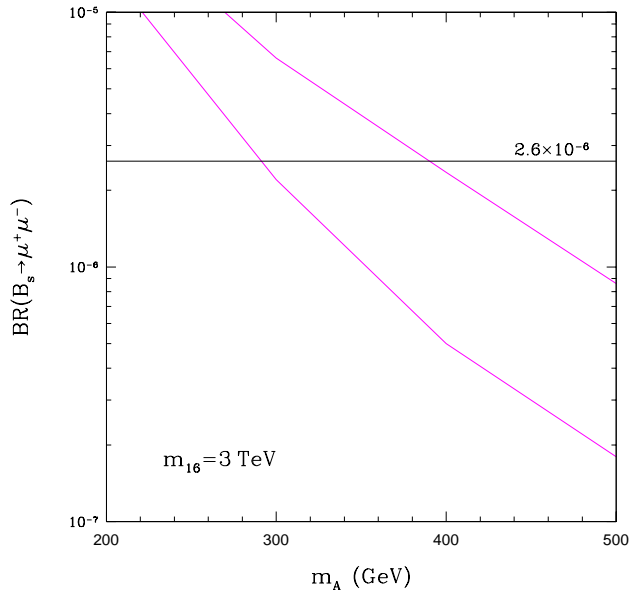,height=3.5in}
\end{minipage}
\caption{\label{fig:bsmm} {\small The upper and lower limits on
${\rm BR} (B_s \rightarrow \mu^+\ \mu^-)$ as a function of $m_{A}$
in the $\mu, \ M_{1/2}$ region of parameter space satisfying all
the collider constraints, $0.095 < \abundchi < 0.13$ and $\chi^2 <
2$ for fixed $m_{16} = 3$ TeV.  } }
\end{center}
\end{figure}

To summarize, we find that regions satisfying all three
constraints exist for $m_{16} \geq 3$ TeV and $m_{A} \geq 300$
GeV.  The acceptable range for $\mu, M_{1/2}$ grows with
increasing $m_{A}$ (for fixed $m_{16} = 3$ TeV) from approximately
$ 260 \; {\rm GeV} \leq \mu \leq 295 \; {\rm GeV}$, $140 \; {\rm
GeV} \leq M_{1/2} \leq 200 \; {\rm GeV}$ for $m_{A} = 300 \; {\rm
GeV}$ to $160 \; {\rm GeV} \leq \mu \leq 520 \; {\rm GeV}$, $140
\; {\rm GeV} \leq M_{1/2} \leq 530 \; {\rm GeV}$ for $m_{A} = 500
\; {\rm GeV}$.  In addition, the allowed regions grow as $m_{16}$
increases and for $m_{A} = 500 \; {\rm GeV},\;\; m_{16} = 5 \;
{\rm TeV}$ there is also a large $\mu, \ M_{1/2}$ solution
satisfying $480 \; {\rm GeV} \leq \mu \leq 820 \; {\rm GeV}, \;
640 \; {\rm GeV} \leq M_{1/2} \leq 1020 \; {\rm GeV}$. In the
Table we present the input and output data from the $\chi^2$
analysis for three points satisfying all the phenomenological
constraints.

\begin{table}
\label{t:table}
$$
\begin{array}
{|l|c|c|c|c|}
\hline
{\rm Data \; points}  & & 1 & 2 & 3   \\
\hline
 {\rm Input\; parameters}  & &  & &  \\
\hline
\;\;\;\alpha_{G}^{-1} & & 24.66 & 24.92  & 25.28  \\
\;\;\; M_G \times 10^{-16} & & 3.51 &  2.83 &  2.43   \\
\;\;\;\epsilon_3 & & -0.038  & -0.034  & -0.029  \\
\;\;\;\lambda  & &  0.66  &  0.66  & 0.66  \\
\hline
\;\;\; m_{16} & & 3000 & 3000  & 5000  \\
\;\;\; m_{10}/m_{16} & & 1.30 & 1.33 & 1.33  \\
\;\;\; \Delta m_H^2  & & 0.14 &  0.15 &  0.14 \\
\;\;\; M_{1/2} & & 180 &  400 &  700  \\
\;\;\;\mu & & 270 &  350 &  600  \\
\;\;\; \tan\beta & & 50.9  &  50.6 &  50.5   \\
\;\;\; A_0/m_{16} &  & -1.85 &  -1.88 &  -1.91   \\
\hline \hline
\chi^2 \; {\rm observables} & {\rm Exp}\;(\sigma)  &    & &   \\
\hline
\;\;\;M_Z & 91.188 \;(0.091) & 91.18 &  91.19 &  91.20  \\
\;\;\;M_W & 80.419 \;(0.080) & 80.42 &  80.42 &  80.41   \\
\;\;\;G_{\mu}\times 10^5 & 1.1664 \;(0.0012) & 1.166 & 1.166 &
1.166 \\
\;\;\;\alpha_{EM}^{-1} & 137.04\; (0.14) & 137.0 & 137.0 &  137.0   \\
\;\;\;\alpha_s(M_Z)  & 0.118 \; (0.002)& 0.1177 &  0.1176 &  0.1179  \\
\;\;\;\rho_{new}\times 10^3 & -0.200 \; (1.10) & 0.427 &  0.498 & 0.162 \\
\hline
\;\;\;M_t   &174.3\; (5.1)    & 173.9 &  174.7 &  174.7  \\
\;\;\;m_b(m_b) & 4.20 \; (0.20)  &  4.28  &  4.28 &  4.21  \\
\;\;\;M_{\tau} & 1.7770 \; ( 0.0018)  & 1.777 &  1.777 &  1.777  \\
\hline
 {\rm TOTAL}\;\;\;\; \chi^2 &  & 0.53 & 0.61 & 0.13  \\
\hline \hline
\;\;\; h      &        &  120  & 119  & 117  \\
\;\;\; H       &        &  329   & 556 & 557 \\
\;\;\; A        &       &  299  & 499  & 501  \\
\;\;\; H^+         &      &  329  & 540 & 541 \\
\;\;\; \chi^0_1 &  &  72  & 163  & 293   \\
\;\;\; \chi^0_2  & &  133  & 288  & 536 \\
\;\;\; \chi^+_1  & &   133  & 287 & 535  \\
\;\;\; \tilde g         &  &  474  & 1032  & 1768  \\
\;\;\; \tilde t_1       &  &  300  & 300  & 576  \\
\;\;\; \tilde b_1       & &  679 & 736 & 1262  \\
\;\;\; \tilde \tau_1    &  & 870 & 721 & 1180  \\
\hline \;\;\; a_\mu^{SUSY} \times 10^{10}  & 25.6 \; (16) & 2.7 &
2.8  & 1.0 \\
\;\;\; \Omega_{\chi}h^2  &  0.095 - 0.130
  & 0.099 & 0.130 & 0.097 \\
\;\;\; \sigsip (pb)\times 10^7  &   & 1.020  & 0.158 &  0.049  \\
\;\;\; {\rm BR} (B_s \rightarrow \mu^+\ \mu^- ) \times 10^6 & <
2.6 & 2.58  & 0.61  & 0.66 \\
\;\;\; {\rm BR} (B \rightarrow X_s \gamma)\times 10^4 & 3.41 \;(0.67)
& 5.36 & 4.34 & 0.81 \\
\hline
\end{array}
$$
\end{table}

\section{ Predictions and Summary} \label{sec:predictions}

In this paper we have analyzed the MSO$_{10}$SM and found regions
of soft SUSY breaking parameter space which fit precision
electroweak data, including the top, bottom and tau masses and, in
addition, fit the cosmological dark matter abundance for the
neutralino LSP and satisfy ${\rm BR} (B_s\rightarrow \mu^+\
\mu^-)$.  Generically, we find solutions to all the constraints
with $m_{16} \geq 3$ TeV.  The squark and slepton masses have an
inverted scalar mass hierarchy with the first and second
generation scalar masses of order $m_{16}$, while the third
generation has mass less than 1.3 TeV for $m_{16} = 5$ TeV.  This
nice feature of the model suppresses SUSY CP and flavor problems.
In addition the gaugino masses are typically much lighter, except
for the large $\mu, \ M_{1/2}$ region for $m_{16} = 5$ TeV with a
gluino mass of order 1.7 TeV (see spectrum in Table for selected
acceptable points).

Note an immediate consequence of such heavy first and second
generation sleptons is the suppression of the SUSY contribution to
the anomalous magnetic moment of the muon.   We find $a_\mu^{SUSY}
\leq 2.8 \times 10^{-10}$ (see Table).  This is consistent with
the most recent experimental~\cite{gminus2} and theoretical
results at $1 \sigma$ if one uses $\tau$--based
analysis~\cite{Davier:2002dy}. However it is only consistent with
an $e^+ e^-$--based analysis at 3 $\sigma$.

Another interesting result is the enhanced branching ratio for the
process $B_s \rightarrow \mu^+\ \mu^-$.  In
Fig.~\protect\ref{fig:bsmm} we show the ranges of values of ${\rm
BR} (B_s \rightarrow \mu^+\ \mu^-)$ in the low $\mu, \ M_{1/2}$
region of parameter space satisfying all the phenomenological
constraints with $0.095 < \abundchi < 0.13$ and $\chi^2 < 2$ as a
function of $m_{A}$ for fixed $m_{16} = 3$ TeV. The horizontal red
line is the CDF bound. Over a significant region of parameter
space ${\rm BR} (B_s \rightarrow \mu^+\ \mu^-) > 1 \times 10^{-7}$
and may be observable at the Tevatron (Run
II)~\cite{Arnowitt:2002cq}.

\begin{figure}[t!]
\begin{center}
\hspace*{-1.5in}
\begin{minipage}{3.5in}
\epsfig{file=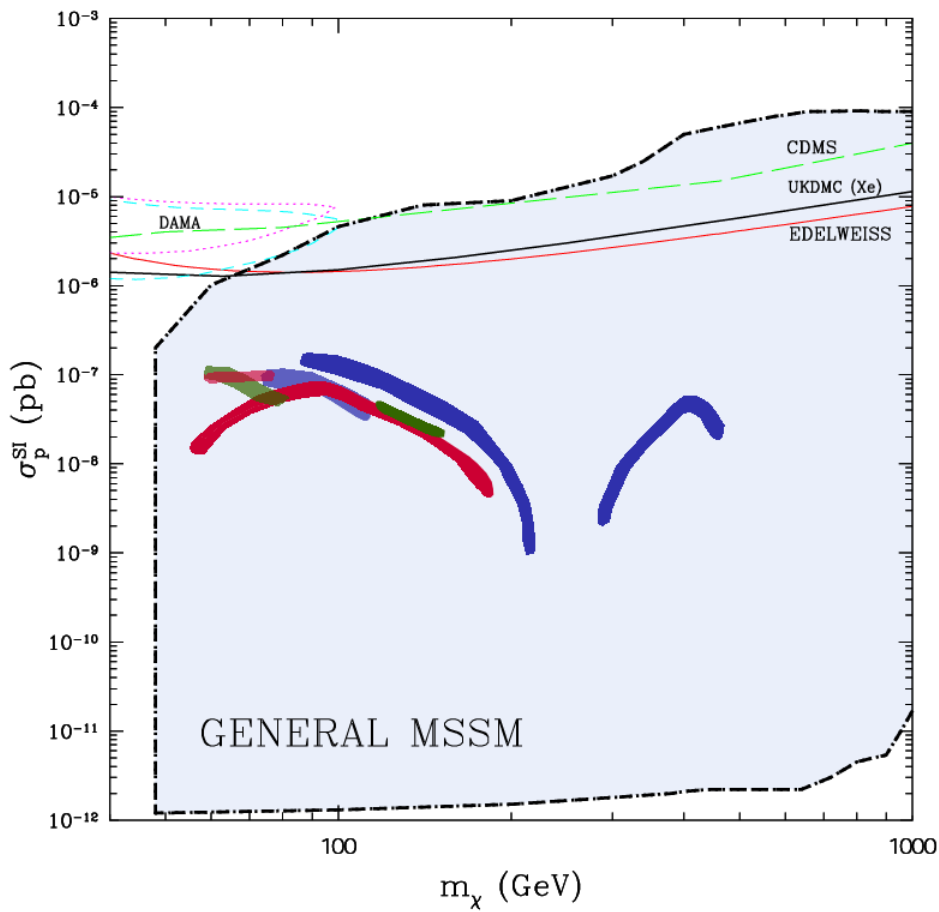,height=5in}
\end{minipage}
\caption{\label{fig:sigsip} {\small Predictions for $\sigsip$
    \vs\ $\mchi$ for different choices of $m_{16}$ and $m_{A}$,
    subject to the collider constraints, $0.095 < \abundchi < 0.13$ and
    $\chi^2<3$.  The green bands are for $m_{16}=2.5$ TeV, the red for
    3 TeV and the blue for 5 TeV. The lighter shading is for
    $m_{A}=300$ GeV, the darker for $m_{A}=500$ GeV. In the last case
    ($m_{16}=5$ TeV and $m_{A}=500$ GeV) there are two branches which
    correspond to the two cosmologically preferred regions in the upper
    right panel in Fig.~\protect\ref{fig:varm0ma}.  } }
\end{center}
\end{figure}

Finally in Fig.~\protect\ref{fig:sigsip} we present the
cross--section for elastic neutralino--proton scattering due to
scalar interactions $\sigsip$ for all regions satisfying the
collider constraints, $0.095 < \abundchi < 0.13$ and $\chi^2<2$.
The green bands are for $m_{16}=2.5$ TeV, the red for 3 TeV and
the blue for 5 TeV. The lighter shading is for $m_{A}=300$ GeV,
the darker for $m_{A}=500$ GeV. In the last case ($m_{16}=5$ TeV
and $m_{A}=500$ GeV) there are two branches which correspond to
the two cosmologically preferred regions in the upper right panel
in Fig.~\protect\ref{fig:varm0ma}.   Note that lower $m_{A}$
generally gives larger $\sigsip$ as expected. For comparison, we
also show the bounds from the present dark matter searches and the
predictions of the general MSSM~\cite{knrr1}. (Other recent
studies of $\sigsip$ in the case of non--universal Higgs mass in a
variant of the CMSSM can be found in~\cite{nuhmstudy}.)  Over the
next two to five years the experimental sensitivity is expected to
gradually improve by some three orders of magnitude. This will
cover large parts of the predicted ranges of $\sigsip$, especially
at lower values of $m_A$ where $B_s \rightarrow \mu^+\ \mu^-$ will
be accessible at the Tevatron (Run~II).

\acknowledgments We gratefully acknowledge the use of the GUT
$\chi^2$ analysis code developed by T. Bla\v{z}ek and a
contribution from Y.G.~Kim to a routine for computing ${\rm BR}
(B_s \rightarrow \mu^+ \ \mu^-)$.  L.R. is grateful to A.
Melchiorri for helpful comments regarding the recent WMAP
analysis. R.D. is supported, in part, by the U.S. Department of
Energy, Contract DE-FG03-91ER-40674 and the Davis Institute for
High Energy Physics.  S.R. received partial support from DOE
grant\# DOE/ER/01545-841 and from a grant in aid from the Monell
Foundation.  He is also grateful to the hospitality shown to him
by the School of Natural Sciences, Institute for Advanced Study.
RRdA and LR are supported in part by the EU Fifth Framework
network "Supersymmetry and the Early Universe"
(HPRN-CT-2000-00152).


\begin{thebibliography}{99}

\bibitem{cmssm}
G.~L.~Kane, C.~F.~Kolda, L.~Roszkowski and J.~D.~Wells,
Phys.\ Rev.\ D {\bf 49}, 6173 (1994) [arXiv:hep-ph/9312272].

\bibitem{bdr}
T.~Blazek, R.~Dermisek and S.~Raby,
Phys.\ Rev.\ Lett.\  {\bf 88}, 111804 (2002)
[arXiv:hep-ph/0107097];
Phys.\ Rev.\ D {\bf 65}, 115004 (2002) [arXiv:hep-ph/0201081].


\bibitem{Tobe:2003bc}
K.~Tobe and J.~D.~Wells,
[arXiv:hep-ph/0301015].

\bibitem{Auto:2003ys}
D.~Auto, H.~Baer, C.~Balazs, A.~Belyaev, J.~Ferrandis and X.~Tata,
[arXiv:hep-ph/0302155].

\bibitem{ewsb}  M.~Olechowski and S.~Pokorski,
Phys.\ Lett.\ B {\bf 344}, 201 (1995) [arXiv:hep-ph/9407404];
D.~Matalliotakis and H.~P.~Nilles,
Nucl.\ Phys.\ B {\bf 435}, 115 (1995) [arXiv:hep-ph/9407251];
N.~Polonsky and A.~Pomarol,
Phys.\ Rev.\ D {\bf 51}, 6532 (1995) [arXiv:hep-ph/9410231];
H.~Murayama, M.~Olechowski and S.~Pokorski,
Phys.\ Lett.\ B {\bf 371}, 57 (1996) [arXiv:hep-ph/9510327];
R.~Rattazzi and U.~Sarid,
Phys.\ Rev.\ D {\bf 53}, 1553 (1996) [arXiv:hep-ph/9505428].

\bibitem{Chattopadhyay:2001va}
U.~Chattopadhyay, A.~Corsetti and P.~Nath,
Phys.\ Rev.\ D {\bf 66}, 035003 (2002) [arXiv:hep-ph/0201001].

\bibitem{Pallis:2003jc}
C.~Pallis and M.~E.~Gomez,
arXiv:hep-ph/0303098.

\bibitem{wmap0302}
See, \eg, latest WMAP results in D.~N.~Spergel {\it et al.},
[arXiv:astro-ph/0302209].


\bibitem{bsmumu}
C.~Hamzaoui, M.~Pospelov and M.~Toharia,
Phys.\ Rev.\ D {\bf 59}, 095005 (1999) [arXiv:hep-ph/9807350];
K.~S.~Babu and C.~F.~Kolda,
Phys.\ Rev.\ Lett.\  {\bf 84}, 228 (2000) [arXiv:hep-ph/9909476];
P.~H.~Chankowski and L.~Slawianowska,
Phys.\ Rev.\ D {\bf 63}, 054012 (2001) [arXiv:hep-ph/0008046];
C.~Bobeth, T.~Ewerth, F.~Kruger and J.~Urban,
Phys.\ Rev.\ D {\bf 64}, 074014 (2001) [arXiv:hep-ph/0104284];
ibid.
Phys.\ Rev.\ D {\bf 66}, 074021 (2002) [arXiv:hep-ph/0204225];
A.~Dedes, H.~K.~Dreiner and U.~Nierste,
Phys.\ Rev.\ Lett.\  {\bf 87}, 251804 (2001)
[arXiv:hep-ph/0108037];
G.~Isidori and A.~Retico,
JHEP {\bf 0111}, 001 (2001) [arXiv:hep-ph/0110121];
A.~J.~Buras, P.~H.~Chankowski, J.~Rosiek and L.~Slawianowska,
Phys.\ Lett.\ B {\bf 546}, 96 (2002) [arXiv:hep-ph/0207241].


\bibitem{cdf}  F.~Abe {\it et al.}  [CDF Collaboration],
Phys.\ Rev.\ D {\bf 57}, 3811 (1998).

\bibitem{Balazs:2003mm}
C.~Balazs and R.~Dermisek,
arXiv:hep-ph/0303161.

\bibitem{scrunching}  J.~A.~Bagger, J.~L.~Feng, N.~Polonsky and R.~J.~Zhang,
Phys.\ Lett.\ B {\bf 473}, 264 (2000) [arXiv:hep-ph/9911255].

\bibitem{masieroetal}  F.~Gabbiani, E.~Gabrielli, A.~Masiero and L.~Silvestrini,
Nucl.\ Phys.\ B {\bf 477}, 321 (1996) [arXiv:hep-ph/9604387];
T.~Besmer, C.~Greub, and T.~Hurth,
\npb{609}, 359 (2001) [arXiv:hep-ph/0105292].

\bibitem{superk}  E.~Kearns, Snowmass 2001, http://hep.bu.edu/

\bibitem{pdecay} R.~Dermisek, A.~Mafi and S.~Raby,
Phys.\ Rev.\ D {\bf 63}, 035001 (2001) [arXiv:hep-ph/0007213].


\bibitem{chi2} T.~Blazek, M.~Carena, S.~Raby and C.~E.~Wagner,
Phys.\ Rev.\ D {\bf 56}, 6919 (1997) [arXiv:hep-ph/9611217].


\bibitem{carenaetal} H.~E.~Haber and R.~Hempfling,
Phys.\ Rev.\ D {\bf 48}, 4280 (1993) [arXiv:hep-ph/9307201];
 M.~Carena, J.~R.~Espinosa, M.~Quiros and C.~E.~Wagner,
Phys.\ Lett.\ B {\bf 355}, 209 (1995) [arXiv:hep-ph/9504316];
M.~Carena, M.~Quiros and C.~E.~Wagner,
Nucl.\ Phys.\ B {\bf 461}, 407 (1996) [arXiv:hep-ph/9508343].


\bibitem{pdg2000}  The Review of Particle Physics, D.E. Groom, et al.,
The European Physical Journal {\bf C15}, 1 (2000).

\bibitem{Beneke:1999fe}
M.~Beneke and A.~Signer,
Phys.\ Lett.\ B {\bf 471}, 233 (1999) [arXiv:hep-ph/9906475].

G.~Corcella and A.~H.~Hoang,
Phys.\ Lett.\ B {\bf 554}, 133 (2003) [arXiv:hep-ph/0212297].

\bibitem{rhonew} P. Langacker, talk at Chicagoland seminar, October (1999).

\bibitem{nrr1+2}
T.~Nihei, L.~Roszkowski and R.~Ruiz de Austri,
JHEP {\bf 0105}, 063 (2001) [arXiv:hep-ph/0102308];
%
JHEP {\bf 0203}, 031 (2002) [arXiv:hep-ph/0202009].

\bibitem{eg97}
J.~Edsjo and P.~Gondolo,
Phys.\ Rev.\ D {\bf 56}, 1879 (1997) [arXiv:hep-ph/9704361].


\bibitem{nrr3}
T.~Nihei, L.~Roszkowski and R.~Ruiz de Austri,
JHEP {\bf 0207}, 024 (2002) [arXiv:hep-ph/0206266].

\bibitem{darksusy}
P.~Gondolo, J.~Edsjo, L.~Bergstrom, P.~Ullio, and T.~Baltz, \\
{\tt http://www.physto.se/edsjo/darksusy/}.

\bibitem{Arnowitt:2002cq}
R.~Arnowitt, B.~Dutta, T.~Kamon and M.~Tanaka,
Phys.\ Lett.\ B {\bf 538}, 121 (2002) [arXiv:hep-ph/0203069].

\bibitem{bsgamma-exp}
S.~Chen, {\it et al.}  [CLEO Collaboration],
Phys. Rev. Lett.  {\bf 87}, 251807 (2001) [arXiv:hep-ex/0108032].
R.~Barate, {\it et al.}  [ALEPH Collaboration],
Phys. Lett. {\bf B429}, 169 (1998);
K.~Abe, {\it et al.}  [Belle Collaboration],
Phys. Lett. {\bf B511}, 151 (2001) [arXiv:hep-ex/0103042];
C.~Jessop, {\it et al.} [BaBar Collaboration], talk at ICHEP--02,
Amsterdam, July, 2002.

\bibitem{or1}
K.~Okumura and L.~Roszkowski,
[arXiv:hep-ph/0208101].

\bibitem{gm01}
P.~Gambino and M.~Misiak,
Nucl. Phys. {\bf B611}, 338 (2001) [arXiv:hep-ph/0104034].

\bibitem{bcmu02}
A.J.~Buras, A.~Czarnecki, M.~Misiak and J.~Urban,
Nucl. Phys. {\bf B631}, 219 (2002) [arXiv:hep-ph/0203135],
and reference therein.

\bibitem{dgg00}
G.~Degrassi,  P.~Gambino and G.~F.~Giudice,
JHEP {\bf 0012}, 009 (2000) [arXiv:hep-ph/0009337].

\bibitem{cgnw00}
M.~Carena, D.~Garcia, U.~Nierste and C.~E.~Wagner,
Phys. Lett. {\bf B499}, 141 (2001) [arXiv:hep-ph/0010003].

\bibitem{rrn1}
L.~Roszkowski, R.~Ruiz de Austri and T.~Nihei,
JHEP {\bf 0108}, 024 (2001) [arXiv:hep-ph/0106334].



\bibitem{gminus2} G.~W.~Bennett {\it et al.}  [Muon g-2 Collaboration],
Phys.\ Rev.\ Lett.\  {\bf 89}, 101804 (2002) [Erratum--ibid.\  {\bf
89}, 129903 (2002)] [arXiv:hep-ex/0208001].


\bibitem{Davier:2002dy}
M.~Davier, S.~Eidelman, A.~Hocker and Z.~Zhang,
[arXiv:hep-ph/0208177].



\bibitem{knrr1} Y.~G.~Kim, T.~Nihei, L.~Roszkowski and R.~Ruiz de Austri,
\jhep{0212}{034}{2002}, 
[arXiv:hep-ph/0208069].




\bibitem{nuhmstudy}
For recent analyses see, \eg, J.~Ellis, A.~Ferstl, K.A.~Olive,
Y.~Santoso,
[arXiv:hep-ph/0302032]; V.~Bertin, E.~Nezri, J.~Orloff,
[arXiv:hep-ph/0210034].




\end{thebibliography}
\end{document}